\documentclass[a4paper]{jpconf}
\usepackage{graphicx}
\usepackage{amsmath}
\usepackage{amssymb}
\usepackage{lineno}
\usepackage[colorlinks,linkcolor=blue,citecolor=blue]{hyperref}

\begin{document}

\title{Neutron polarisation correction to triple-axis data with analytical derivations}

\author{Y Nambu$^{1,2,3}$, M Enderle$^4$, T. Weber$^4$ and K Kakurai$^5$}
\address{$^1$ Institute for Materials Research, Tohoku University, Sendai 980-8577, Japan}
\address{$^2$ Organization for Advanced Studies, Tohoku University, Sendai 980-8577, Japan}
\address{$^3$ FOREST, Japan Science and Technology Agency, Kawaguchi, Saitama 332-0012, Japan}
\address{$^4$ Institut Laue-Langevin, 38042 Grenoble, France}
\address{$^5$ Neutron Science and Technology Center, Comprehensive Research Organization for Science and Society, Tokai, Ibaraki 319-1106, Japan}

\ead{nambu@tohoku.ac.jp}

\begin{abstract}
    Polarised neutron scattering is the method of choice to study magnetism in condensed matter.
    Polarised neutrons are typically very low in flux, and complex experimental configurations further reduce the count rate.
	Neutron polarisation corrections would therefore be needed.
    Here we analytically derive formulae of the corrected partial differential scattering cross-sections.
    The analytical method is designed for the longitudinal polarisation analysis, and the correction generally holds for time-independent polarised neutrons with a triple-axis spectrometer.
    We then apply the correction to recent results of our $P_x$ experiment on Y$_3$Fe$_5$O$_{12}$.
    Although there is a difficulty with the experimental determination of inefficiency parameters of neutron spin polarisers and flippers, the correction appears to work properly.
\end{abstract}

Neutron scattering is a powerful microscopic measurement tool for studying condensed matter~\cite{Lovesey1986}.
Scattered neutrons from nuclei and unpaired electron spins, can respectively provide information on the crystal structure and magnetism.
One of the strengths of the neutron scattering technique is that the magnetic scattering cross-section is typically comparable to that of nuclear scattering.
These scattering processes can generally be understood by the linear response theory, enabling quantitative discussion through the measured intensity.

Polarised neutron scattering--taking into account the neutrons' spin degree of freedom in the scattering process--gives further detailed information on the magnetism~\cite{Chatterji2006}.
It has mainly been used to separate the magnetic and nuclear scattering cross-sections~\cite{Moon1969}.
The disentanglement of the magnetic moment directions, as well as the symmetry of magnetic fluctuations~\cite{Kakurai1984}, can also be demonstrated.
More recently the ``chiral term''~\cite{Maleyev1995} was used to measure chiral magnetic order~\cite{Loire2011} and excitations in paramagnetic~\cite{Roessli2002} and chiral phases~\cite{Lorenzo2007a}.
The nuclear-magnetic ``interference term'' also attracts much attention recently.
The nuclear-magnetic interference is possible when nuclear and magnetic scattering occurs at the same positions in reciprocal space.
Examples are found in $\vec{q}_{\rm m}=0$ magnetic structures with $\vec{q}_{\rm m}$ being the magnetic wavevector, and also in structures with $\vec{q}_{\rm m}\ne 0$ in which magnetic order is coupled to a structural distortion with the same $\vec{q}_{\rm m}$ in the context of, for instance, magnetoelastic coupling~\cite{Pramanick2013,Finger2010}.

\begin{figure}[t]
	\begin{center}
		\includegraphics[bb=0 0 445 162,width=\textwidth]{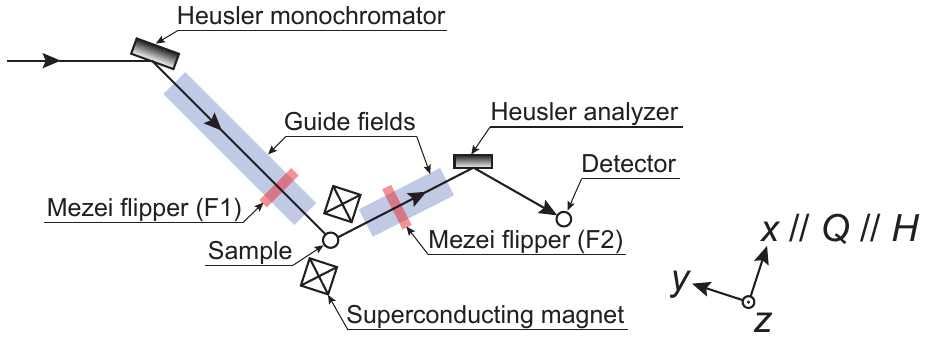}
	\end{center}
	\caption{\label{fig1}Sketch of the $P_x$ experiment on the IN20 instrument with bold black arrows denoting the neutron path. The scattering coordinate $(x,y,z)$ is also given.}
\end{figure}

For schematic understanding, we adopt a simple scattering coordinate $(x,y,z)$ as in Fig.~\ref{fig1}, where $x\parallel \vec{Q}$, $y\perp \vec{Q}$ where $\vec{Q}$ is the scattering wavevector, and $z$ is perpendicular to the horizontal scattering plane.
Observed cross-sections are summarised by the following formulae~\cite{Blume1963,Maleyev1963} within the longitudinal polarisation analysis (LPA),
\begin{align}
	&\sigma_x^{\pm\pm}\propto N,\label{xnsf}\\
	&\sigma_x^{\pm\mp}\propto M_y + M_z \mp M_{\rm ch},\label{xsf}\\
	&\sigma_y^{\pm\pm}\propto N+M_y\pm R_y,\label{ynsf}\\
	&\sigma_y^{\pm\mp}\propto M_z,\label{ysf}\\
	&\sigma_z^{\pm\pm}\propto N+M_z\pm R_z,\label{znsf}\\
	&\sigma_z^{\pm\mp}\propto M_y,\label{zsf}
\end{align}
where the ideal case with perfect neutron spin polarisers and flippers is anticipated, and cross-sections from isotope-incoherent and nuclear-spin incoherent scattering are omitted for simplicity.
The term $\sigma_{\alpha}^{io}$ ($\alpha = x,y,z$) stands for the partial differential scattering cross-section (${\rm d}^2\sigma/({\rm d}\Omega{\rm d}\omega)^{io}$) with $i$ in-coming and $o$ outgoing neutrons with $+/-$ neutron polarisation.
The non-magnetic nuclear ($N=\langle N_{Q}N_{Q}^{\dagger }\rangle _{\omega}$), in-plane magnetic ($M_y=\langle M_{Qy}M_{Qy}^{\dagger}\rangle _{\omega}$) and out-of-plane magnetic ($M_z=\langle M_{Qz}M_{Qz}^{\dagger}\rangle _{\omega}$), chiral ($M_{\rm ch}=i(\langle M_{Qy}M_{Qz}^{\dagger}\rangle_{\omega}-\langle M_{Qz}M_{Qy}^{\dagger}\rangle_{\omega})$), and nuclear-magnetic interference ($R_{\beta}=\langle N_Q M_{Q\beta}^{\dagger}\rangle_{\omega}+\langle M_{Q\beta}N_Q^{\dagger}\rangle_{\omega}$) ($\beta = y,z$) terms are included.
$\langle N_{Q}N_{Q}^{\dagger }\rangle _{\omega }$ and $\langle M_{Q\beta}M_{Q\beta}^{\dagger}\rangle_{\omega}$ are the spatiotemporal Fourier transforms of the nuclear-nuclear and spin-spin correlation functions, respectively.
The chiral term defines the antisymmetric correlation function within the $yz$-plane, and the interference terms describe the symmetric part of the nuclear-magnetic interference.

In reality, measured intensities can, however, easily be contaminated by not-perfectly-working neutron spin polarisers and flippers, and result in linear combinations within the four intrinsic cross-sections; $\sigma^{++}$, $\sigma^{+-}$, $\sigma^{-+}$, and $\sigma^{--}$.
Polarised neutrons typically have very low flux, and the special scattering geometry required for the cases of within scattering plane neutron polarisations, $P_x$ and $P_y$, further reduce the count rate.
Thus neutron polarisation corrections are occasionally needed, and their conventional techniques are summarised~\cite{Williams1988}.
However, these techniques have appeared not to be universally applicable and will collapse in certain experimental conditions.
To overcome this, a universal method of correcting polarisation data is proposed~\cite{Wildes1999,Wildes2006}, being rigorously valid for one-dimensional LPA.
Here we provide analytically derived forms of the corrected cross-sections and demonstrate an application of it to our recent $P_x$ experiment~\cite{Nambu2020}.

We here raise the first observation of the magnon mode-resolved direction of the precessional motion of the magnetic order, i.e., magnon polarisation~\cite{Nambu2020,Nambu2021}.
In spintronics with insulating ordered magnets, the spin current--a flow of the spin degree of freedom--can be carried by the transverse component of the spin wave.
Given that this transverse component is mathematically equivalent to the magnon polarisation, information on the sign of the magnon polarisation and its quantitative elucidation are necessary for understanding the microscopic mechanism of the spin current.
The magnon polarisation is known to affect thermodynamic properties of the spin current, including the magnitude and sign of the spin Seebeck effect.
We thus performed the observation using the garnet Y$_3$Fe$_5$O$_{12}$ (YIG).

YIG is a ferrimagnetic insulator, and is nowadays quintessential for microwave and optical technologies, and fundamental research in spintronics, magnonics, and quantum information.
To detect the magnon polarisation, neutron spins shall be aligned parallel or antiparallel to $\vec{Q}$.
Magnetic neutron scattering can only detect the spin components perpendicular to $\vec{Q}$, and these projections are tiny due to an application of the magnetic field ($\vec{H}\parallel \vec{Q}$).
The component detected by this configuration corresponds to the area that the precessional motion covers, and the use of polarised neutrons can distinguish clockwise and anticlockwise motion through the chiral term detection ($M_{\rm ch}$ in eq.~(\ref{xsf})).
The chiral term is typically employed to observe a spatial variation of the {\it non-collinear} magnetic moments caused by effects such as geometrical frustration and/or Dzyaloshinskii--Moriya interactions.
Here we aim for a separate intrinsic property in a {\it collinear} magnet.

Inelastic polarised neutron scattering data~\cite{data} were collected on the thermal neutron triple-axis spectrometer IN20~\cite{IN20} at the Institut Laue-Langevin, France.
We used a graphite filter in the outgoing beam to suppress higher-order contaminations and fixed the final wavenumber of $k_{\rm f}=2.662$~{\AA}$^{-1}$, corresponding to final energies of $E_{\rm f}=14.7$~meV.
No collimator was put into the beam path to maximise the neutron flux.
A YIG single crystal ($\sim 8$~g) was oriented with $(HHL)$ in the horizontal scattering plane, and inserted into a cryomagnet supplying a horizontal magnetic field ($\vec{H}\parallel x$) of 0.3~T and temperatures between 10~K and 300~K.
The obtained fields are homogeneous over a large area wrapping the sample position and continuously connecting to the guide fields along the neutron path.
The magnetic field direction was aligned parallel to $\vec{Q}$.
YIG is a very soft magnet~\cite{Yamasaki2009}, and an external field of 0.3~T is sufficient to fully saturate the magnetisation into a single magnetic domain.

As depicted in Fig.~\ref{fig1}, the polarised neutron triple-axis spectrometer IN20~\cite{IN20} consists of the Heusler monochromator (polariser), two Mezei spin flippers, and Heusler analyser.
The monochromator and analyser can reflect a beam with only spin-polarised neutrons.
We define the upper stream flipper as F1 and the downstream as F2.
Flippers rotate the neutron spin 180$^{\circ}$ so that the spin state after the flipper is antiparallel to the state before.
It is assumed that a flipper does not affect the beam when it is not used.
IN20, like other all polarised neutron instruments, requires guide fields in the neutron flight path to maintain the neutron polarisation.

In the experiment, the polarised neutron scattering intensity is recorded in four {\it channels}: with both flippers off $I_{00}$ ($=I_{x}^{--}$), with F1 on and F2 off $I_{10}$ ($=I_{x}^{+-}$), vice-versa $I_{01}$ ($=I_{x}^{-+}$), and with both flippers on $I_{11}$ ($=I_{x}^{++}$).
The four partial differential scattering cross-sections, $\sigma^{++}$, $\sigma^{+-}$, $\sigma^{-+}$, and $\sigma^{--}$, can be extracted after the polarisation correction.
Following the work by A.~R.~Wildes~\cite{Wildes1999}, the relation between $I$ and $\sigma$ can be expressed in a matrix form,
\begin{align}
	\begin{pmatrix}
		I_{00}\\
		I_{01}\\
		I_{10}\\
		I_{11}
	\end{pmatrix}
	=&
	\begin{pmatrix}
		1&0&0&0\\
		0&1&0&0\\
		f_1&0&1-f_1&0\\
		0&f_1&0&1-f_1
	\end{pmatrix}
	\begin{pmatrix}
		1&0&0&0\\
		f_2&1-f_2&0&0\\
		0&0&1&0\\
		0&0&f_2&1-f_2
	\end{pmatrix}
	\nonumber\\
	&\times
	\begin{pmatrix}
		1-p_1&0&p_1&0\\
		0&1-p_1&0&p_1\\
		p_1&0&1-p_1&0\\
		0&p_1&0&1-p_1
	\end{pmatrix}
	\begin{pmatrix}
		1-p_2&p_2&0&0\\
		p_2&1-p_2&0&0\\
		0&0&1-p_2&p_2\\
		0&0&p_2&1-p_2
	\end{pmatrix}
	\begin{pmatrix}
		\sigma^{--}\\
		\sigma^{-+}\\
		\sigma^{+-}\\
		\sigma^{++}
	\end{pmatrix}.
	\label{eq-wildes}
\end{align}

The efficiency of the polarising and flipping elements could be imperfect, and the measured intensity will therefore be linear combinations of the four $\sigma^{io}$.
Equation~(\ref{eq-wildes}) describes the relation in terms of $p_1$ (Heusler monochromator), $p_2$ (Heusler analyser), $f_1$ (F1), and $f_2$ (F2), where $p_i$ and $f_i$ ($i =1,2$) denote the probability of neutrons not being perfectly polarised and flipped, respectively.
An unpolarised neutron beam is defined with $p_i=0.5$, and the case for $p_i=0$ corresponds to the perfectly polarised beam.
Likewise, $f_i$ can take 0 if the flipper works perfectly.

To extract the partial differential scattering cross-sections $\sigma^{io}$, the inverse matrices can be solved analytically,
\begin{align}
	\sigma^{--}=&
	I_{00}\left[\frac{f_1}{1-f_1}\left\{\frac{p_1(1-p_2)}{(1-2p_1)(1-2p_2)}+\frac{f_2 p_1 p_2}{(1-f_2)(1-2p_1)(1-2p_2)}\right\}\right.\nonumber\\
	&\qquad \left.+\frac{(1-p_1)(1-p_2)}{(1-2p_1)(1-2p_2)}
	+\frac{f_2(1-p_1)p_2}{(1-f_2)(1-2p_1)(1-2p_2)}\right]
	\nonumber\\
	&-I_{01}\left\{\frac{(1-p_1)p_2}{(1-f_2)(1-2p_1)(1-2p_2)}+\frac{f_1 p_1 p_2}{(1-f_1)(1-f_2)(1-2p_1)(1-2p_2)}\right\}
	\nonumber\\
	&-I_{10}\left[\frac{1}{1-f_1}\left\{\frac{p_1(1-p_2)}{(1-2p_1)(1-2p_2)}+\frac{f_2 p_1 p_2}{(1-f_2)(1-2p_1)(1-2p_2)}\right\}\right]
	\nonumber\\
	&+I_{11}\frac{p_1 p_2}{(1-f_1)(1-f_2)(1-2p_1)(1-2p_2)},
	\end{align}
	\begin{align}
	\sigma^{-+}=&
	-I_{00}\left[\frac{f_1}{1-f_1}\left\{\frac{p_1 p_2}{(1-2p_1)(1-2p_2)}+\frac{f_2 p_1(1-p_2)}{(1-f_2)(1-2p_1)(1-2p_2)}\right\}\right.
	\nonumber\\
	&\qquad \left.+\frac{(1-p_1)p_2}{(1-2p_1)(1-2p_2)}+\frac{f_2(1-p_1)(1-p_2)}{(1-f_2)(1-2p_1)(1-2p_2)}\right]
	\nonumber\\
	&+I_{01}\left\{\frac{(1-p_1)(1-p_2)}{(1-f_2)(1-2p_1)(1-2p_2)}+\frac{f_1 p_1(1-p_2)}{(1-f_1)(1-f_2)(1-2p_1)(1-2p_2)}\right\}
	\nonumber\\
	&+I_{10}\left[\frac{1}{1-f_1}\left\{\frac{p_1 p_2}{(1-2p_1)(1-2p_2)}+\frac{f_2 p_1(1-p_2)}{(1-f_2)(1-2p_1)(1-2p_2)}\right\}\right]
	\nonumber\\
	&-I_{11}\frac{p_1(1-p_2)}{(1-f_1)(1-f_2)(1-2p_1)(1-2p_2)},
\end{align}
\begin{align}
	\sigma^{+-}=&
	-I_{00}\left[\frac{f_1}{1-f_1}\left\{\frac{(1-p_1)(1-p_2)}{(1-2p_1)(1-2p_2)}+\frac{f_2(1-p_1)p_2}{(1-f_2)(1-2p_1)(1-2p_2)}\right\}\right.
	\nonumber\\
	&\qquad \left.+\frac{p_1(1-p_2)}{(1-2p_1)(1-2p_2)}+\frac{f_2 p_1 p_2}{(1-f_2)(1-2p_1)(1-2p_2)}\right]
	\nonumber\\
	&+I_{01}\left\{\frac{p_1 p_2}{(1-f_2)(1-2p_1)(1-2p_2)}+\frac{f_1(1-p_1)p_2}{(1-f_1)(1-f_2)(1-2p_1)(1-2p_2)}\right\}
	\nonumber\\
	&+I_{10}\left[\frac{1}{1-f_1}\left\{\frac{(1-p_1)(1-p_2)}{(1-2p_1)(1-2p_2)}+\frac{f_2(1-p_1)p_2}{(1-f_2)(1-2p_1)(1-2p_2)}\right\}\right]
	\nonumber\\
	&-I_{11}\frac{(1-p_1)p_2}{(1-f_1)(1-f_2)(1-2p_1)(1-2p_2)},
\end{align}
\begin{align}
	\sigma^{++}=&
	I_{00}\left[\frac{f_1}{1-f_1}\left\{\frac{(1-p_1)p_2}{(1-2p_1)(1-2p_2)}+\frac{f_2(1-p_1)(1-p_2)}{(1-f_2)(1-2p_1)(1-2p_2)}\right\}\right.
	\nonumber\\
	&\qquad \left.+\frac{p_1 p_2}{(1-2p_1)(1-2p_2)}+\frac{f_2 p_1(1-p_2)}{(1-f_2)(1-2p_1)(1-2p_2)}\right]
	\nonumber\\
	&-I_{01}\left\{\frac{p_1(1-p_2)}{(1-f_2)(1-2p_1)(1-2p_2)}+\frac{f_1(1-p_1)(1-p_2)}{(1-f_1)(1-f_2)(1-2p_1)(1-2p_2)}\right\}
	\nonumber\\
	&-I_{10}\left[\frac{1}{1-f_1}\left\{\frac{(1-p_1)p_2}{(1-2p_1)(1-2p_2)}+\frac{f_2(1-p_1)(1-p_2)}{(1-f_2)(1-2p_1)(1-2p_2)}\right\}\right]
	\nonumber\\
	&+I_{11}\frac{(1-p_1)(1-p_2)}{(1-f_1)(1-f_2)(1-2p_1)(1-2p_2)}.
\end{align}
These represent the general relations between the measured intensity and intrinsic cross-sections $\sigma^{io}$, and quantitative discussion is eased by using the inefficiency parameters $p_i$ and $f_i$ if these can be estimated.
An example of the estimation of the inefficiencies of instruments is provided in Ref.~\cite{Wildes2006}.

Given that $p_1$, $p_2$, $f_1$, and $f_2$ can be energy dependent, it is not trivial to experimentally extract the inefficiency parameters.
In addition, for horizontal field experiments using $P_x$ or $P_y$ neutron polarisations, $f_1$ and the polarisation transport in the incident wavenumber $k_{\rm i}$ depend on the angle between the magnetic field and $k_{\rm i}$.
Likewise, $f_2$ and the polarisation transport in the final wavenumber $k_{\rm f}$ depend on the angle between the magnetic field and $k_{\rm f}$.
All parameters depend on the field strength.

We here consider the case of a purely nuclear scatterer with $\sigma^{++}=\sigma^{--}$ and $\sigma^{+-}=\sigma^{-+}=0$, $f_1$, $f_2$ and the average polarisation $\phi \equiv \left(1-2p_1\right)\left(1-2p_2\right)$ can be derived as,
\begin{align}
	f_1&=\frac{I_{00}-I_{01}+I_{10}+I_{11}}{2\left(I_{00}-I_{01}\right)},\\
	f_2&=\frac{I_{00}+I_{01}-I_{10}+I_{11}}{2\left(I_{00}-I_{10}\right)},\\
	\phi&=\frac{\left(I_{00}-I_{01}\right)\left(I_{00}-I_{10}\right)}{I_{00}I_{11}-I_{01}I_{10}}.
\end{align}
We measured $I_{00}$, $I_{01}$, and $I_{10}$ channels using the direct beam at the elastic $k_{\rm i}=k_{\rm f}=2.662$~\AA$^{-1}$ with applying magnetic field of 0.3~T, as a function of the angle between the magnetic field and the $k_{\rm i}=k_{\rm f}$ direction.
Measured intensities are summarised in Table~\ref{table}.

\begin{center}
	\begin{table}[t]
	\caption{\label{table}Measured intensity using the direct beam.}
	\centering
	\begin{tabular}{cccc}
	\br
	Angle (deg) & $I_{00}$ & $I_{01}$ & $I_{10}$\\
	\mr
	30 & 3910 & 251.5 & 290.0 \\
	50 & 3932 & 274.8 & 255.4 \\
	70 & 4048 & 296.2 & 262.4 \\
	80 & 4086 & 304.8 & 255.4 \\
	90 & 4043 & 296.4 & 277.2 \\
	100 & 3877 & 290.5 & 257.6 \\
	110 & 4001 & 288.9 & 261.1 \\
	120 & 3915 & 291.8 & 252.4 \\
	130 & 3923 & 275.8 & 269.7 \\
	140 & 3869 & 273.0 & 254.5 \\
	150 & 3748 & 262.9 & 255.5 \\
	\br
	\end{tabular}
	\end{table}
\end{center}

Using the above formulae, we derive the following relations,
\begin{align}
	\frac{F_1}{F_2}\equiv\frac{1-f_1}{1-f_2}&=\frac{I_{00}-I_{10}}{I_{00}-I_{01}},\\
	F_1\frac{2\phi}{1+\phi}&=1-\frac{I_{10}}{I_{00}},\\
	F_2\frac{2\phi}{1+\phi}&=1-\frac{I_{01}}{I_{00}}.
\end{align}
As calculated from Table~\ref{table}, the statistical error ($\sqrt{I_{00}}$) of the $I_{00}$ count rates is of the order of 1.6\%, and the fluctuations in $F_1/F_2$ against the angle are about 1\%.
We hereafter assume that neither $\phi$ nor $F_1$ nor $F_2$ depend on the angle between the magnetic field and $k_{\rm i}$ or $k_{\rm f}$ in the tested angular range.
On average, $F_1$ is 0.5\% smaller than $F_2$.
Considering the statistical error of these measurements, we may assume $F_2\sim 0.9953$ and $F_1\sim 0.990$, leading to $\phi=0.8827$.
The average $\phi$ of the direct beam cannot be smaller than 0.8749 since $F_2 \le 1$.
Our experimental setup, however, did not allow to separate $p_1$ and $p_2$, nor to separate $f_1$ and $f_2$ from the polarising and analysing efficiency and the polarisation transport across the sample~\cite{Yerozolimsky1999}.
On the direct beam with $k_{\rm i} = k_{\rm f} = 2.662$ {\AA}$^{-1}$, flipping ratios range from 13 to 16 for each flipper (Table~\ref{table}).
This limits the worst-case beam polarisation of the spectrometer with the horizontal magnetic field, where we presume the perfectly working flippers with $f_1=f_2=0$, yielding $0.030<p_1=p_2<0.037$.
If we set the beam polarisation as good as $p_1 = p_2 = 0.019$, which corresponds to 93\% neutron polarisation, we obtain the worst-case flipper inefficiencies, namely 0.025 to 0.04.
For further precise estimations, a chiral scatterer with $\sigma^{++}=\sigma^{--}$ and $\sigma^{+-}\ne\sigma^{-+}$ would allow to determine $\left(1-2p_1\right)/\left(1-2p_2\right)$.
Together with pre-determined $\phi=\left(1-2p_1\right)\left(1-2p_2\right)$, $p_1$ and $p_2$ can be determined separately.

\begin{figure}[t]
	\begin{center}
		\includegraphics[bb=0 0 2644 2149,width=\textwidth]{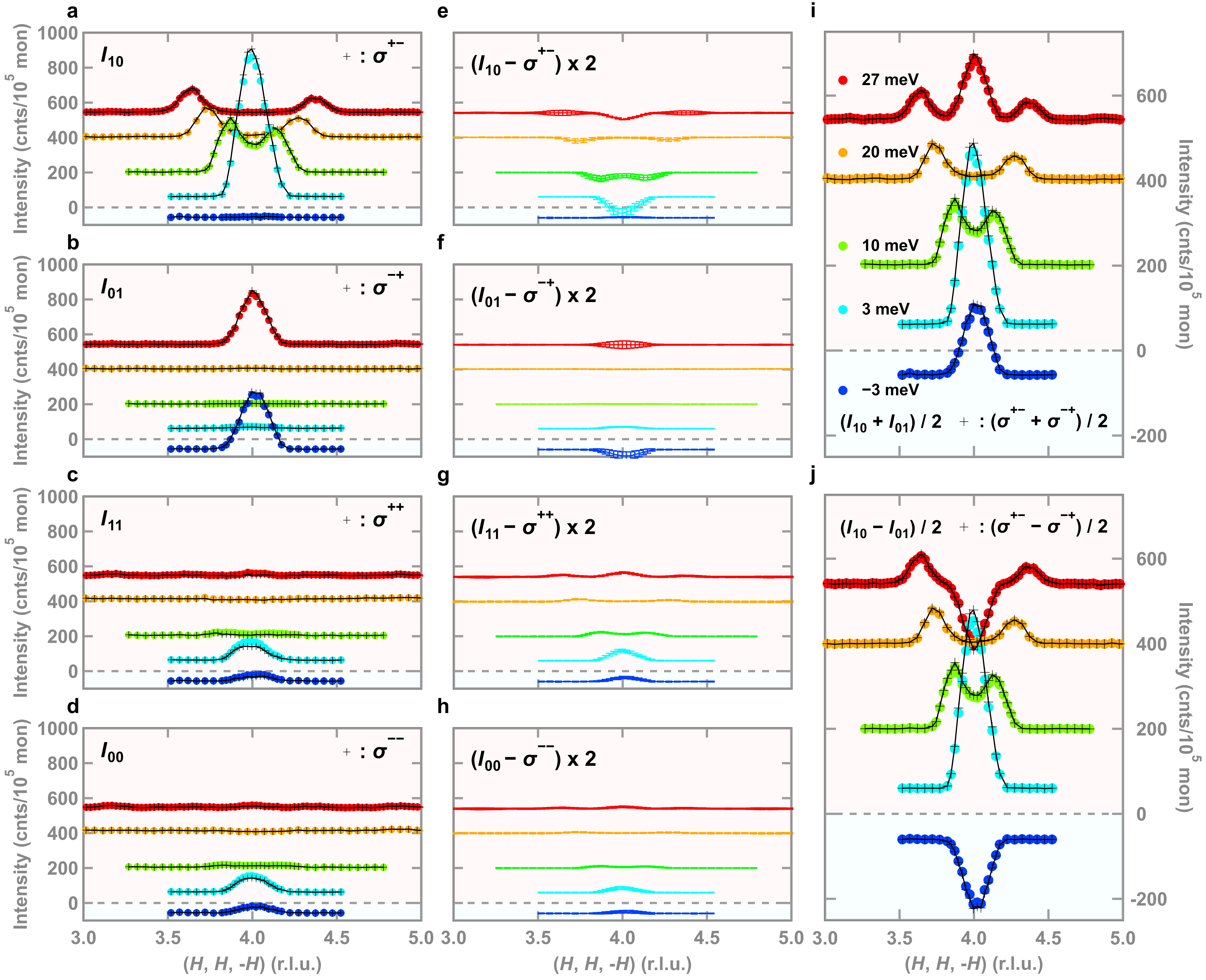}
	\end{center}
	\caption{\label{fig2}Uncorrected raw data from IN20~\cite{Nambu2020} (filled circle) and corrected data (black cross) are compared. Constant energy scans of (a) $I_{10}$, (b) $I_{01}$, (c) $I_{11}$, and (d) $I_{00}$ channels at representative energy transfers. Note that spectra are shifted vertically for $20\times E$, where $E$ denotes the energy transfer. The differences between uncorrected and corrected data are emphasised by doubled intensities in (e), (f), (g), and (h). Derived intensity of (i) the magnetic: $M=M_y+M_z\propto\frac{1}{2}(I_{10}+I_{01})$ and (f) chiral term: $M_{\rm ch}\propto\frac{1}{2}(I_{10}-I_{01})$ compared with corrected data. All the scans run along the P$[11\bar{1}]$ direction and were taken at the fixed final wavenumber $k_{\rm f}=2.662$ {\AA}$^{-1}$.}
\end{figure}

We here assume $p_1 = p_2 = 0.019$ and $f_1 = f_2 = 0.025$.
Using these approximated inefficiency parameters, Fig.~\ref{fig2} compares uncorrected raw data taken at 293~K~\cite{Nambu2020} and corrected partial differential scattering cross-sections.
Opposite magnon polarisation is already apparent without any correction, the polarisation correction indeed works for the partial differential scattering cross-sections $\sigma^{++}$ (Fig.~\ref{fig2}(c)) and $\sigma^{--}$ (Fig.~\ref{fig2}(d)).
Reflecting the weak nuclear contribution around the $(4,4,-4)$ reflection, the corrected data are suppressed compared to the uncorrected data.
This can be illustrated in the differences between uncorrected and corrected data plotted in Fig.~\ref{fig2}(e-h).
Such small values of $p_i$, $f_i$ and resultant slight changes between the uncorrected and corrected data, underline the nearly {\it perfect} experimental performance of IN20 in the LPA mode.
If the inefficiency parameters $f_1$, $f_2$, $p_1$, and $p_2$ are negligibly small like in IN20, only spin-flip channels can extract the chiral term.
However, this ensures only qualitative discussion, and all four channels are required for quantitative discussion with entirely derived corrected data.
One could perform a half-polarised mode with the incident or scattered neutron polarisation, whose formulae are
\begin{align}
	&\sigma_x^{+0}=\sigma_x^{++}+\sigma_x^{+-}\nonumber\\
	=\ &\sigma_x^{0-}=\sigma_x^{+-}+\sigma_x^{--}\propto N+M_y+M_z-M_{\rm ch},\\
	&\sigma_x^{-0}=\sigma_x^{-+}+\sigma_x^{--}\nonumber\\
	=\ &\sigma_x^{0+}=\sigma_x^{++}+\sigma_x^{-+}\propto N+M_y+M_z+M_{\rm ch}.
\end{align} 
The main advantage of the half-polarised mode is that either a polariser or analyser is required, which enhances neutron intensity compared to the full polarisation mode.
However, the half-polarised mode gives less capability to disentangle nuclear and magnetic terms, $M_y$ and $M_z$ in this particular case.

To summarise, we derived inverse matrices to calculate the corrected partial differential scattering cross-sections with the inefficiency
parameters of neutron spin polarisers and flippers.
These formulae generally hold for time-independent polarised neutrons with a triple-axis spectrometer, and they can thus be widely applied to LPA on the polarised neutron scattering data.
We treated recent results of our $P_x$ experiment on YIG as an example, and confirmed that the correction successfully works.
An accurate estimate of the inefficiency parameters will be a future issue.

\ack
We thank M.~B{\"o}hm for his assistance during the experiment.
This work was partly supported by Grants-In-Aid for Scientific Research (Grants No.~16H04007, No.~17H06137, No.~21H03732, and No.~22H05145) from MEXT of Japan, FOREST (Grant No.~JPMJFR202V) from JST, and the Graduate Program in Spintronics at Tohoku University.

\end{document}